\documentstyle[12pt,./psfig,./aaspp4]{article}

\newcommand{\Wo}{\mbox{${\rm W}_0$}}
\newcommand{\msun}{\mbox{${\rm M}_\odot$}}

\newcommand{\SeBa}{\mbox{${\sf SeBa}$}}

\newcommand{\nbody}{\mbox{{{\em N}-body}}}
\newcommand{\thm}{\mbox{${t_{\rm hm}}$}}

\newcommand{\trxh}{\mbox{${t_{\rm rxh}}$}}
\newcommand{\trxt}{\mbox{${t_{\rm rxt}}$}}

\newcommand{\mgal}{\mbox{${M_{\rm Gal}}$}}

\newcommand{\rcore}{\mbox{${r_{\rm core}}$}}

\newcommand{\rhm}{\mbox{${r_{\rm hm}}$}}
\newcommand{\rgc}{\mbox{${r_{\rm GC}}$}}

\newcommand{\rtide}{\mbox{${r_{\rm tide}}$}}

\newcommand{\rLf}{\mbox{${r_{\rm L1}}$}}

\def\unit#1{{\mbox{[{\rm #1}]}}}
\def\apgt{\ {\raise-.5ex\hbox{$\buildrel>\over\sim$}}\ }
\def\aplt{\ {\raise-.5ex\hbox{$\buildrel<\over\sim$}}\ }

\lefthead{S.\ F.\ Portegies Zwart et al.}
\righthead{Young star clusters near the Galactic center}

\begin{document}


\title{How many young star clusters exist in the Galactic center?}

\medskip 

\author{Simon F.\ Portegies Zwart,$^1$
	Junichiro Makino,$^2$
	Stephen L.\ W.\ McMillan,$^3$
	and
	Piet Hut$^4$
       }

\vfill\noindent
$^1$ Massachusetts Institute of Technology, Cambridge, MA 02139, USA,
{\em Hubble Fellow} \\
$^2$ Department of Astronomy, University of Tokyo, 7-3-1 Hongo,
     Bunkyo-ku,Tokyo 113-0033, Japan \\
$^3$ Dept.\ of Physics,
		  Drexel University, 
                  Philadelphia, PA 19104, USA \\
$^4$ Institute for Advanced Study, 
		Princeton, NJ 08540, USA \\
\bigskip


Subject headings: gravitation --- methods: n-body simulations
		--- stellar dynamics --- stars: evolution 
		--- globular clusters: individual (Arches, Quintuplet)

\newpage

\begin{abstract}\noindent
We study the evolution and observability of young compact star
clusters within $\sim200$\,pc of the Galactic center.  Calculations
are performed using direct {\nbody} integration on the GRAPE-4,
including the effects of both stellar and binary evolution and the
external influence of the Galaxy.  The results of these detailed
calculations are used to calibrate a simplified model applicable over
a wider range of cluster initial conditions.
We find that clusters within 200\,pc of the Galactic center dissolve
within $\sim70$ Myr.  However, their projected densities drop below
the background density in the direction of the Galactic center within
$\sim 20$\,Myr, effectively making these clusters undetectable after
that time.  Clusters farther from the Galactic center but at the same
projected distance are more strongly affected by this selection
effect, and may go undetected for their entire lifetimes.
Based on these findings, we conclude that the region within 200 pc of
the Galactic center could easily harbor some 50 clusters with
properties similar to those of the Arches or the Quintuplet systems.

\end{abstract}

\section{Introduction}
Two young compact star clusters have been observed within a few tens
of parsecs of the Galactic center: the Arches cluster (Object 17,
Nagata et al.\, 1995) and the Quintuplet cluster (AFGL\,2004, Nagata
et al.\, 1990; Okuda et al.\, 1990), for which excellent observational
data are available.  In terms of structural parameters---size, mass,
density profile---and ages, these systems may represent the Galactic
counterparts to NGC\,2070 (R\,136), the central star cluster in the
30\,Doradus region in the Large Magellanic Could (Massey \& Hunter
1998).  The Arches and Quintuplet clusters lie behind thick layers of
absorbing material, hinting that many more such systems may exist.
Recently, Dutra \& Bica (2000) have reported from the 2MASS survey a
total of 58 star cluster candidates within $\sim600$\,pc (in
projection) of the Galactic center.

A number of important questions make these clusters worthy of detailed
study, among them: (1) How are such clusters related to globular
clusters?  (2) How do they contribute to the total star formation rate
in the Galaxy?  (3) Are their mass functions in reality intrinsically
flat, as is suggested by observations?  (4) How far are these clusters
 from the Galactic center?  (5) How many are hidden, still waiting to
be discovered?  In this Letter we summarize the results of a series of
{\nbody} simulations of young compact star clusters in the vicinity of
the Galactic center, and address specifically the last item on this
list.

We found that the lifetimes of our model clusters depend very
sensitively on their distances from the Galactic center.  This is
mainly due to the larger size of tidally limited clusters lying
farther from the Galactic center, resulting in longer relaxation times
and therefore longer lifetimes.  The majority of our models are
visible only for the first part of their lifetimes, and are likely to
be indistinguishable from the stellar background at later times.  We
find that the true number of young compact star clusters within 200 pc
of the Galactic center is at least 10 but could easily exceed 50.

A more comprehensive paper, exploring all of these questions in more
detail and presenting the results of an extensive parameter-space
study, is in preparation (Portegies Zwart et al., 2000b).

%
\section{Initial conditions}
We study the evolution of our model clusters by following the
equations of motion of all stars by direct {\nbody} integration, at
the same time taking into account the internal evolution of both stars
and binary systems.  The ``Starlab'' software environment within which
this work was performed is described in detail by Portegies Zwart et
al.\, (2000a; see also {\tt http://www.sns.ias.edu/$\sim$starlab}).
The special-purpose GRAPE-4 system (Makino et al.\ 1997) was used to
accelerate the computation of gravitational forces between stars.

%

Observed parameters for the Arches and Quintuplet clusters are listed
in Table\,\ref{Tab:observed}.  These clusters have masses of $\sim
10^4$\msun\, and are extremely compact, with half-mass radii
$\rhm\aplt 1$\,pc (Figer, McLean \& Morris 1999).  The projected
distance from the Arches to the Galactic center is about 34\,pc; the
Quintuplet cluster lies somewhat farther out, at $\sim$50\,pc.

%
%

\begin{table*}[ht]
\caption[]{Observed  parameters  for  the  Arches and  the  Quintuplet
clusters.  Columns list cluster  name, reference, age, mass, projected
distance  to the Galactic  center, tidal radius ($\rtide$),  and half
mass radius  ($\rhm$).  The final  column presents an estimate  of the
density within the half mass radius.  }
\begin{flushleft}
\begin{tabular}{ll|rrrrrr} \hline
Name 	  &ref& Age  &   M     & \rgc & \rtide & \rhm   
				& $\rho_{\rm hm}$ \\ 
          &&[Myr]& [$10^3$\,\msun] & \multicolumn{3}{c}{------ [pc] ------}   &
			 [$10^5$\,\msun/pc$^2$] \\ \hline
Arches    &a& 1--2 & 12--50   & 30   & 1    &  0.2     & 0.6  -- 2.6 \\ 
Quintuplet&b& 3--5 & 10--16   & 50   & 1    &$\sim 0.5$& 0.08 -- 0.13 \\
\hline
\end{tabular} \\
\smallskip
References:
a) Brandl et al.\,(1996);
   Campbell et al.\,(1992);
   Massey \& Hunter (1998).
b) Figer et al.\,(1999);
\end{flushleft}
\label{Tab:observed} 
\end{table*}

Our calculations start with 12k (12288) stars at zero age.  We choose
stellar masses between 0.1\,{\msun} and 100\,{\msun}, distributed
according to the mass function suggested for the Solar neighborhood by
Scalo (1986).  The median mass of this mass function is about
0.3\,\msun; the mean mass is $\langle m \rangle \simeq 0.6\,\msun$.
The initial mass of each model is therefore $\sim 7500$\,\msun. 
Initially all stars are single, although some binaries do form
dynamically via three-body encounters, in which one star carries away
sufficient energy and angular momentum to allow two others to become
bound.  We adopt three standard distances from the Galactic center,
34\,pc, 90\,pc and 150\,pc.  The initial density profiles and velocity
dispersions for our models are generated from anisotropic Heggie \&
Ramamani (1995) models with $\Wo = 4$.  At birth, the clusters are
assumed to precisely fill their Jacobi surfaces
(``Roche lobes'') in the tidal field of the Galaxy, and are taken to
move in circular orbits around the Galactic center.  For a circular
orbit in the plane of the Galaxy the distance from the center of the
star cluster to the first Lagrangian point (the Jacobi radius) is
approximated by
\begin{equation}
	\rLf \simeq \left( {M \over 2\mgal(\rgc)} 
	      \right)^{1/3} \rgc
\label{Eq:L1}\end{equation}
Here $M$ is the mass of the star cluster and the factor two is a
correction factor which depends on the the density profile; strictly
speaking the factor of two is correct only in the case $\mgal \propto
r$.  (The Jacobi radius is computed consistently with the adopted
tidal field in our simulations.)  Table\,\ref{Tab:N12kinit} presents
an overview of our model initial conditions.

\begin{table*}[ht]
\caption[]{Overview of initial conditions for our model calculations.
 Each calculation is performed three times.
 From left to right the columns list the model name, the distance to
 the Galactic center, the initial King parameter \Wo, the initial
 tidal-- and half mass relaxation times, half-mass crossing time, core
 radius, half-mass radius and distance to the first Lagrangian point
 in the tidal field of the Galaxy, the time of core collapse, and the
 time at which the cluster mass drops below 1\% of its initial
 value. 
 }\begin{flushleft}
\begin{tabular}{lrr|rrrrll|rr} \hline
Model  &$r_{\rm gc}$
            &\Wo~&\trxt&\trxh& \thm&\rcore&\rhm & \rLf   &$t_{cc}$ 
						      &$t_{\rm end}$ \\
       &[pc]&   &\multicolumn{2}{c}{---[Myr]---}&[kyr]
	&\multicolumn{3}{c|}{------ [pc] ------}
	&\multicolumn{2}{c}{---[Myr]---} \\ 
\hline
R34W4  & 34 & 4& 53&  3.2 &  27 & 0.05 & 0.117& 0.77  &0.8& 12.7  \\
R90W4  & 90 & 4&134&  8.1 &  68 & 0.09 & 0.218& 1.42  &1.2& 32.6  \\ 
R150W4 &150 & 4&218& 13   & 110 & 0.14 & 0.301& 1.97  &2.0& 53.4  \\ 
\hline     		       
\end{tabular}
\end{flushleft}
\label{Tab:N12kinit} \end{table*}


The mass of the Galaxy within the clusters' orbit at a distance {\rgc}
($\aplt200$\,pc) is taken to be (Mezger et al.\,
1999)\nocite{1999A&A...348..457M}
\begin{equation}
	\mgal(\rgc) = 4.25 \times 10^6 \left({\rgc \over \unit{pc}}
		     	           \right)^{1.2}           \;\;[\msun].
\label{Eq:Mgal}\end{equation}
This mass distribution determines the strength and geometry of the
local Galactic tidal field (for details see Portegies Zwart et al.\,
2000b).  The evolution of the cluster is followed using the Starlab
{\tt kira} {\nbody} integrator and the {\SeBa} binary evolution
program (Portegies Zwart et al.\, 2000b).  For each selected
distance to the Galactic center we carried out three calculations. One
series of runs was carried out with identical initial
realizations of the {\nbody} system (stellar masses, positions and
velocities, with a total mass of 7432\,\msun).  The same initial model
can be used at several galactocentric distances because the
shape of the zero-velocity surface does not depend sensitively on
distance to the Galactic center.  Table\,\ref{Tab:N12kinit} gives
results of these calculations.  For each galactocentric distance we
also performed two
additional calculations (for a total of 9 runs) with
different initial
realizations of the {\nbody} systems.  These calculations were performed
to study the uncertainties in cluster lifetimes, and to ascertain the
reproducilility of our
results.  The calculations with different initial
realizations produced roughly 10\% spreads in core collapse times
($t_{cc}$) and cluster lifetimes ($t_{\rm
end}$).  For reasons of economy, stars were removed from all {\nbody}
calculations when they exceeded a distance of 3\,{\rLf} from the cluster
center.

\section{Results}
\begin{figure}[htbp!]
\hspace*{1.cm}
\psfig{figure=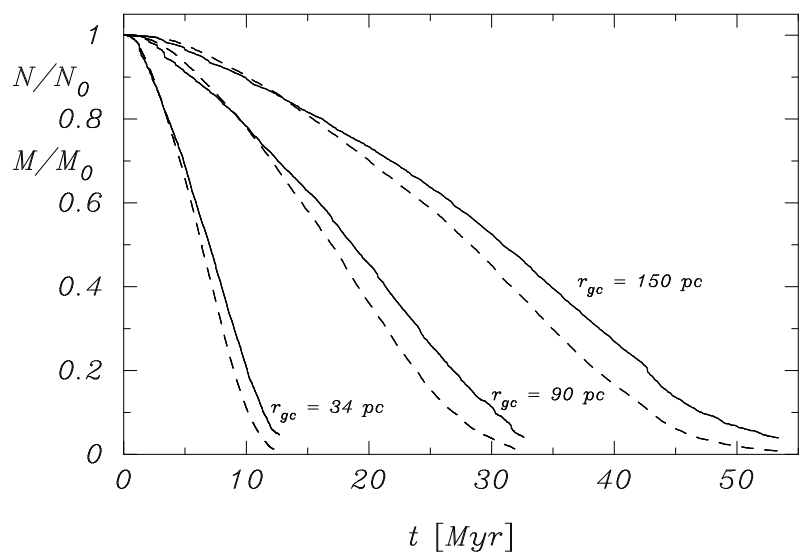,width=12.cm,angle=-90}
\caption[]{Evolution of the total mass $M$ and number of stars $N$
(solid and dashed lines respectively) within the critical
zero-velocity surface of selected models at $\rgc=34$\,pc (left),
$\rgc=90$\,pc (middle) and $\rgc=150$\,pc (right).  
The results of the runs with identical initial realizations are presented.
All quantities are normalized to their initial values.  }
\label{fig:SPZ_W147N12k_TM}
\end{figure}

Figure\,\ref{fig:SPZ_W147N12k_TM} shows the evolution of cluster mass
and number of stars for the models listed in Table
\ref{Tab:N12kinit} which began with identical initial conditions but
different galactocentric distances.
Perhaps not surprisingly, clusters located farther
from the Galactic center live considerably longer than those closer
in.  The longer lifetime of the more distant cluster is mainly a
consequence of its longer relaxation time.  Scaling the relaxation
time at the tidal radius of model R34W4 to a distance of 150\,pc
results in a lifetime of $\sim 52.2$\,Myr ($\equiv 12.7 {\rm Myr}
\times 218/53$), which is slightly smaller than the $\sim 53.4$\,Myr
lifetime of the models R150W4.  
Mass loss from stellar evolution, which
is more prominent in model R150W4 because the cluster survives longer,
seems to be a minor factor in driving the evolution.

The number of stars in each model (dashed lines, see Figure\,1)
decreases more rapidly than the total mass (solid lines).  Thus the
mean mass of the stars within the cluster increases gradually with
time.


\section{Discussion}
Although clusters like the Arches and Quintuplet systems are very
compact, it may still be hard to see them near the Galactic center
because the projected foreground and background stellar density is so
high.  Integration of the local stellar density (obtained by
differentiating Eq.\,\ref{Eq:Mgal}) along the line of sight then gives
the surface density.  Portegies Zwart et al.\,(2000b) perform this
calculation numerically and arrive at a surface density at 34 pc of
about 3000\,\msun\,pc$^{-2}$.  While the contrast in the surface
density of the cluster relative to that of the background is perhaps
an oversimplified measure of the cluster's observability, the results
of this simple comparison are quite instructive, and a more
comprehensive consideration of luminosity density leads to essentially
similar overall conclusions.

\begin{figure}[htbp!]
\hspace*{1.cm}
\psfig{figure=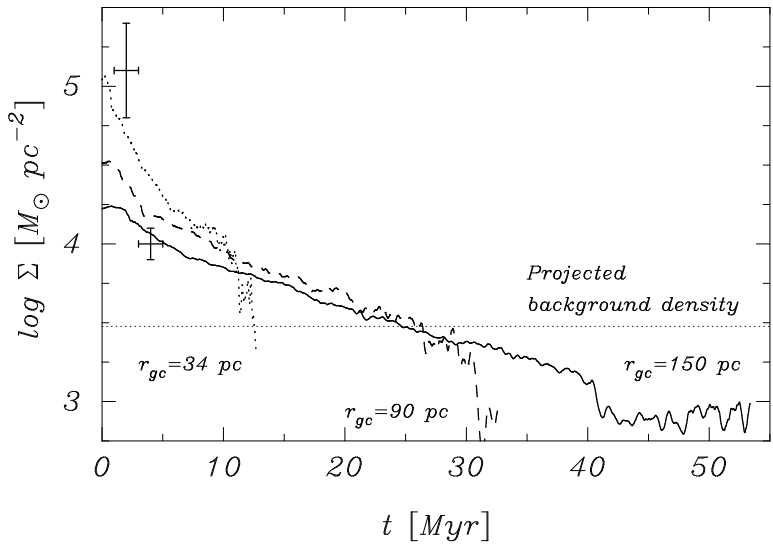,width=12.cm,angle=-90}
\caption[]{ Evolution of the surface density within the projected
half-mass radius for the models at $\rgc = 34$\,pc (dotted line), at
$\rgc = 90$\,pc (dashed line) and at $\rgc = 150$\,pc (solid line).  The
horizontal dotted line gives the integrated background density at a
projected distance of 34\,pc from the Galactic center.  The results of
the runs with identical initial realizations are presented.  The two
error bars give the observed surface densities of the Arches (left)
and the Quintuplet (right) clusters.  }
\label{fig:SPZ_W17R34_rho_phm}
\end{figure}

Figure\,\ref{fig:SPZ_W17R34_rho_phm} shows the evolution of the
surface density within the projected half mass radius for models R34W4
(dots), R90W4 (dashes) and R150W4 (solid).  The two points with error
bars indicate the projected half-mass densities for the Arches (left)
and Quintuplet (right) clusters.  The horizontal dotted line gives the
background surface density at a projected distance of 34\,pc from the
Galactic center.  The projected densities of the two observed clusters
are between 3 (for the Quintuplet) and 50 (for Arches) times higher
than the background; clusters with densities below a few times the
projected background stellar density may well remain unnoticed.

\subsection{A simple model}

A simplified model for the evolution of these star clusters may be
constructed as follows.  The initial relaxation time at the tidal
radius may be calculated using Eq.\,\ref{Eq:Mgal} and Spitzer's (1987)
expression as
\begin{equation}
	\trxt \simeq 2.19 \left( {\rgc \over \unit{pc}} 
		          \right)^{0.9}\; \; \unit{Myr}.
\label{Eq:trlx}\end{equation}
The constant is obtained by substitution of the appropriate units.


The mass of the cluster decreases almost linearly in time (see
Figure\,\ref{fig:SPZ_W147N12k_TM}) as
\begin{equation}
	M = M_0 \left(1 - \frac{\tau}{\tau_c}\right)\,.
\label{Eq:mass}\end{equation}
Here $\tau \equiv t/\trxt$ and $\tau_c \simeq 0.29\trxt$ is the age at which
the cluster dissolves in the tidal field of the Galaxy.  The projected
surface density within the half-mass radius of the cluster is
\begin{equation}
	\Sigma \equiv { M_{\rm hm} \over \pi r^2_{\rm hm} } 
		= { M_{\rm hm} \over \pi w_o^2 r^2_{\rm L1}}.
\label{Eq:Sigma}\end{equation}
Here, $M_{\rm hm} \simeq 0.65M$ is the mass contained within the
projected half-mass radius, and $w_o \equiv \rhm/\rLf$ depends on
the density profile, but is always smaller than unity.  For a King
model with $\Wo = 4$, we find $w_o \simeq 0.16$.  Substitution of
Eqs.\,\ref{Eq:L1} and \ref{Eq:Mgal} into Eq.\,\ref{Eq:Sigma} gives
\begin{equation}
	\Sigma_0 \simeq 7.0 \times 10^6 \left( {\rgc \over \unit{pc}} 
			   \right)^{-1.2} \; \; [\msun\,{\rm pc}^{-2}],
\end{equation}


The projected surface density $\Sigma$ decreases with time because the
cluster mass decreases and the half-mass radius of the cluster
increases. Substitution of Eq.\,\ref{Eq:mass} into Eq.\,\ref{Eq:Sigma}
gives:
\begin{equation}
  \frac{\Sigma(t)}{\Sigma_0} = \left(1 - \frac{\tau}{\tau_c} \right)
			 \ \left(\frac{r_{\rm hm,0}}{\rhm} \right)^2.
\end{equation}
The half-mass radius {\rhm} increases by about a factor of two during
the first half mass relaxation time and remains roughly constant
thereafter.  In our simple model we implement this by allowing $\rhm
\approx r_{\rm hm,0}$ to increase by a factor two in the first half
mass relaxation time and to remain constant at later times.
The fact that the half-mass radius remains roughly constant at late
times, and does not decrease as $M^{1/3}$ as would be expected for a
tidally limited system, is a consequence of our use of the total
{\nbody} mass, rather than the mass within the Jacobi surface (see
Figure\,\ref{fig:SPZ_W147N12k_TM}), in determining both {\rhm} and
$\Sigma$.
The resulting surface density evolution agrees very well with our
{\nbody} calculations.

\begin{figure}[htbp!]
\hspace*{1.cm}


\psfig{figure=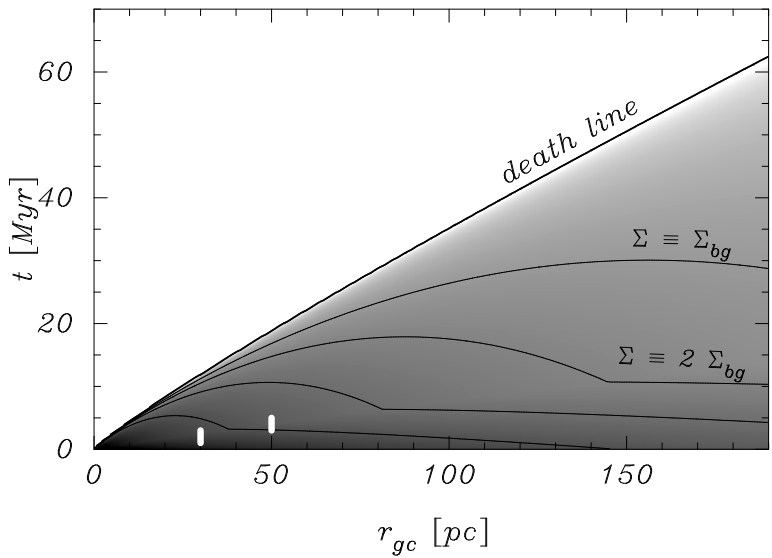,width=12.cm,angle=-90}
\caption[]{Projected cluster density as a function of distance from
the Galactic center and time.  Gray shading indicates projected
surface density; darker shades indicate higher density.  Solid lines
indicate where the surface density of the cluster equals 1, 2, 4, and
10 times the background density.  The background surface density is
computed for 34 pc. The two holes to the lower left indicate the
locations of the Arches (left) and Quintuplet (right) clusters on the
figure.
}
\label{fig:RgcT_dens}
\end{figure}

Figure\,\ref{fig:RgcT_dens} shows the evolution of the projected
density as a function of distance from the Galactic center for our
simple model.  Clusters close to the Galactic center are compact
enough to be easily visible (dark shades) for a large fraction of
their lifetimes, but they dissolve quickly.  Clusters farther from the
Galactic center live much longer, but their surface densities are
generally low, lying well below the background for most of their
lifetimes.  The two well-observed clusters both lie in the lower left
corner (high density region) of Figure\,\ref{fig:RgcT_dens}.  The Arches
and Quintuplet clusters inhabit only a small portion of the available
parameter space.  However, the region they populate is most favorable
for finding clusters, because the projected surface densities of such
clusters are high.

We estimate the number of clusters like the Arches and the Quintuplet
based
on the fact that both clusters lie at the lower left corner (youngest
and most compact) in Figure\,\ref{fig:RgcT_dens}.  We assume
that all clusters with ages less than $\sim5$\,Myr and within 50\,pc
of the
Galactic center have been found by observers, and all outside this
region are as yet undetected.  We further assume that clusters
populate
the triangular area in Figure\,\ref{fig:RgcT_dens} more or less
uniformly---probably not a bad assumption, since the Galactic mass is
roughly proportional to radius (see Equation\,\ref{Eq:Mgal}). With these
assumptions, the number of yet-to-be-found clusters is ($\frac{1}{2}
\times $60\,Myr$\times$200\,pc)/(5\,Myr$\times$50\,pc)=24 times more
than the number of known clusters. Thus, we expect that the total
number of clusters in this region would be around 50.

A conservative lower limit to the number of hidden clusters may be
obtained using the same technique, but adopting the observed projected
density of the Quintuplet system as the limiting contrast at which
such clusters can be discovered. The projected density of the
Quintuplet exceeds the background density by about a factor three.
Figure\,\ref{fig:RgcT_dens} then suggests that about 20\% of the
available surface area harbors visible cluster. A lower limit to the
total number of clusters within 200\,pc of the Galactic center would
then be around 10.

These estimates are in excellent agreement
with the results of Dutra \& Bica (2000) mentioned earlier, although
the above reasoning suggests that even these 58 candidate clusters may
represent only a small fraction of the number actually present.


Finally, we note that a population of 100 clusters with masses of
$10^4$\,{\msun} each and a maximum lifetime of $10^8$\,Myr implies a
star formation rate of 0.01\,{\msun} per year, enough to build up the
entire bulge of $10^8$\,{\msun} within the 10\,Gyr age of the Galaxy.

\acknowledgements 

SPZ is grateful to the Institute for Advanced Study, Drexel University
and Tokyo University for their hospitality and the use of their
GRAPE-4 hardware.  This work was supported by NASA through Hubble
Fellowship grant HF-01112.01-98A awarded by the Space Telescope
Science Institute, which is operated by the Association of
Universities for Research in Astronomy, by the Research for the Future
Program of Japan Society for the Promotion of Science
(JSPS-RFTP97P01102) and by NASA ATP grants NAG5-6964 and NAG5-9264.
Part of this letter was written while SPZ, SM and PH were visiting the
American Museum of Natural History.  They acknowledge the hospitality
of their astrophysics department and visualization group.



\end{document}